\documentclass[10pt]{iopart}

\usepackage{graphicx}
\expandafter\let\csname equation*\endcsname\relax
\expandafter\let\csname endequation*\endcsname\relax
\usepackage{amsmath}
\usepackage{amssymb}
\usepackage{dcolumn}
\usepackage{color}
\usepackage{multirow}
\usepackage{bm}
\usepackage{subfigure}
\usepackage{booktabs}

\begin{document}

\title{\textit{Ab initio} calculations of spin-nonconserving exciton-phonon scattering in monolayer transition metal dichalcogenides }

\author{Xiao-Wei Zhang}
\address{Department of Materials Science and Engineering, University of Washington, Seattle, WA 98195, USA}

\author{Ting Cao}
\address{Department of Materials Science and Engineering, University of Washington, Seattle, WA 98195, USA}

\ead{tingcao@uw.edu}
\vspace{10pt}
\begin{indented}
\item[]December 2021
\end{indented}

\begin{abstract}
%
We investigate the spin-nonconserving relaxation channel of excitons by their couplings with phonons in two-dimensional transition metal dichalcogenides using $\textit{ab initio}$ approaches.
Combining $\text{GW}$-Bethe-Salpeter equation method and density functional perturbation theory, we calculate the electron-phonon and exciton-phonon coupling matrix elements for the spin-flip scattering in monolayer WSe$_{\text{2}}$, and further analyze the microscopic mechanisms influencing these scattering strengths. 
We find that phonons could produce effective in-plane magnetic fields which flip spin of excitons, giving rise to relaxation channels complimentary to the spin-conserving relaxation.
Finally, we calculate temperature-dependent spin-flip exciton-phonon relaxation times. 
Our method and analysis can be generalized to study other two-dimensional materials and would stimulate experimental measurements of spin-flip exciton relaxation dynamics. 
\end{abstract}

%
\vspace{2pc}
%
%
%
\ioptwocol
%

\section{Introduction}
The ability to manipulate, maintain, and detect the spin of electronic quasiparticles not only has the potential to revolutionize traditional electronics, but also holds great promise in many novel applications, such as quantum computing and quantum information science~\cite{vzutic2004spintronics,wolf2006spintronics,hirohata2020review,loss1998quantum,awschalom2018quantum}.
One central focus of spintronics research is to understand the interplay between the spin degree of freedom and the environment where spin-carrying quasiparticles live in, which may result in shorter spin lifetimes and dephasing period ~\cite{vzutic2004spintronics}. 
The interactions between the spin-carrying quasiparticles and phonon bath, namely the spin-phonon scattering, is one contributing factor for the spin-environment coupling.
The spin-phonon scattering, an intrinsic process in materials, plays a vital role in spin relaxation, especially dominant at ambient temperatures~\cite{vzutic2004spintronics}.

In recent years, two-dimensional (2D) materials such as monolayer transition metal dichalcogenides (TMD) are emerging as a fertile platform for studies of spin- and valley-tronics thanks to the existence of strong spin-orbit coupling (SOC) that locks to the emergent valley degree of freedom  ~\cite{xiao2012coupled,cao2012valley, xiao2017excitons,manzeli20172d,wang2018colloquium,xu2014spin,mak2018light,ahn20202d,schaibley2016valleytronics,avsar2020colloquium}.
Furthermore, in contrast to conventional bulk semiconductors, these 2D semiconductors have large exciton binding energies in the order of hundreds of meV due to the reduced dielectric screening and quantum confinement effects that are absent in the bulk ~\cite{wang2018colloquium}. 
As such, these excitons have been considered to host robust valley-spin indices that can encode information.
Nevertheless, spin-flip processes are often found in exciton relaxation and recombination dynamics for monolayer TMD~\cite{li2019emerging,tang2019long}.
It is therefore intriguing to study spin-flip exciton relaxation dynamics in monolayer TMD, which would quantitatively assess the robustness of the excitonic valley-spin indices.
Theoretical calculations of spin-phonon scattering from $\textit{ab initio}$ have made important progress in recent years~\cite{heid2010effect,restrepo2012full,giustino2017electron,xu2020spin,zhang2021phonon}.
However, incorporating the two-body excitonic interactions in phonon-mediated spin relaxation calculations is largely unexplored and challenging~\cite{antonius2017theory}.
The reason lies in the difficulties to accurately calculate the exciton states and the electron-phonon coupling (EPC) matrix elements.
The former problem can be solved by the Bethe-Salpeter equation (BSE) approach~\cite{strinati1988application} with very dense sampling of the Brillouin zone (BZ), especially for 2D materials~\cite{qiu2013optical,qiu2016screening}.
Since phonons can contribute a large quasi-momentum, the finite-momentum BSE is also needed to describe the final exciton states, despite that the initial states, i.e., the optically excited excitons, have zero momentum.
In addition, SOC is crucial to correctly describe phonon-assisted spin-flip scattering.
These factors further increase the complexity of the problem.
Conceptually, it is also important to understand the microscopic mechanisms of spin-flip relaxation since phonon doesn't explicitly carry a spin index.

In this article, we apply the theoretical method implemented previously by the authors~\cite{zhang2021phonon} to investigate the spin-flip exciton relaxation dynamics in 2D semiconductors. 
We calculate the EPC and exciton-phonon matrix element (ExPC) matrix elements fully based on $\textit{ab inito}$ method, in which SOC has been incorporated in the calculations to allow phonon flipping of electron spin.
By analyzing the phonon-mode dependence of EPC matrix elements, we obtain the conditions of spin-flip scattering. 
Finally, we give the intra-valley and inter-valley exciton-phonon scattering times.

\section{Theory and calculation details}
\begin{figure*}[t]
\centering
		\includegraphics[width=0.8\linewidth]{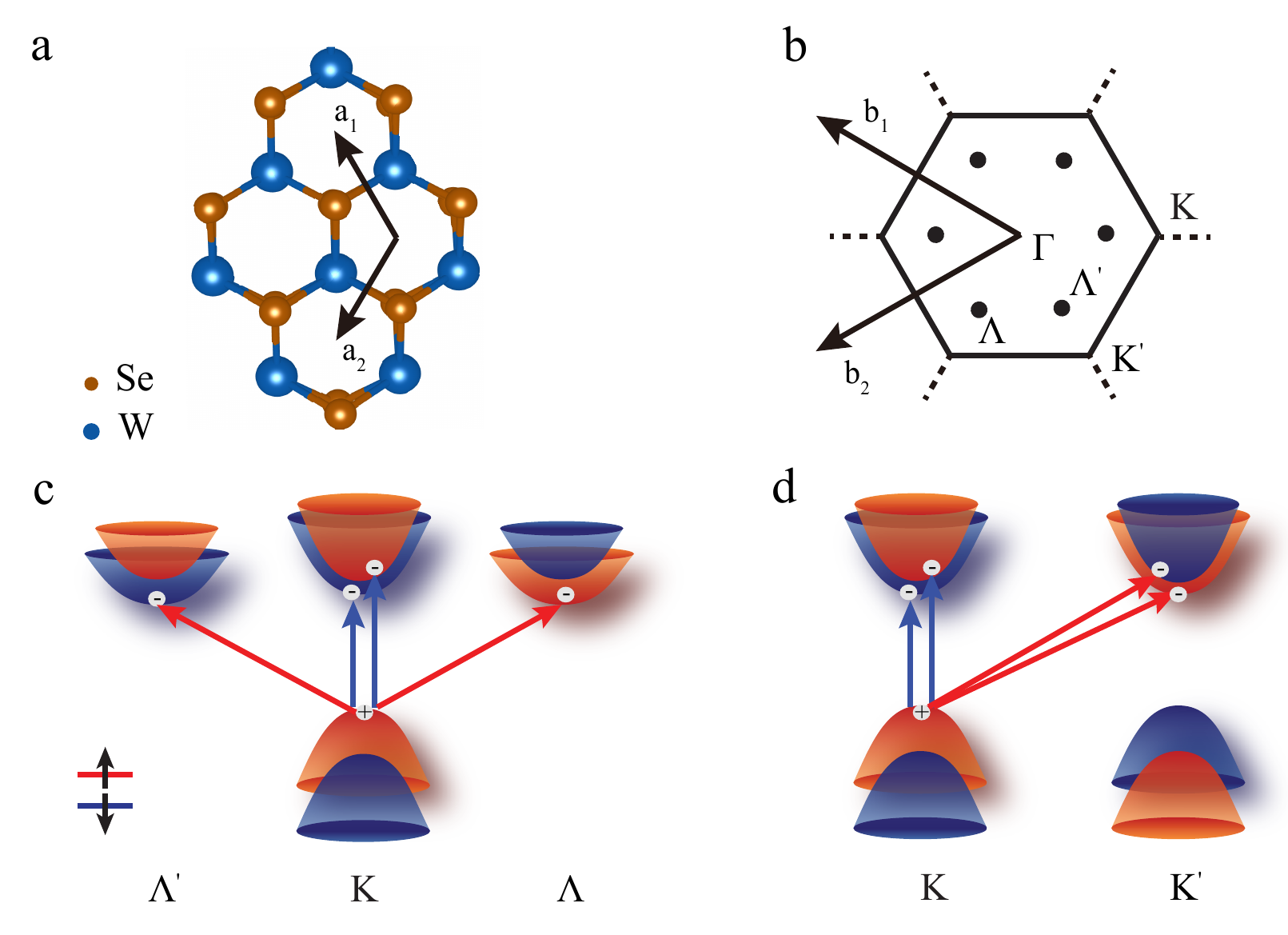}
		\caption{(a) Schematic crystal structure of monolayer WSe$_{\text{2}}$ in top view. (b) The first Brillouin zone corresponding to the unit cell in (a). (c)-(d) Low-energy intra-valley and inter-valley excitons, having the hole in the valence band valleys and the bound electron in the conduction band valleys. The orange and blue colors denote the spin-up and spin-down bands, respectively. 
		}
		\label{fig1}
	\end{figure*}

For the completeness of this work, we summarize the general formalism of exciton-phonon coupling in the current section. Adopting the second-quantization representation, the exciton-phonon coupled Hamiltonian to the lowest order is written as ~\cite{knox1963}:
\begin{align}
H=&\sum_{S\bm{Q}}\Omega_{S\bm{Q}}a_{S\bm{Q}}^{\dagger}a_{S\bm{Q}}+\sum_{\bm{q}\nu}\left(b_{\bm{q}\nu}^{\dagger}b_{\bm{q}\nu}+\frac{1}{2}\right)\hbar\omega_{\bm{q}\nu} \nonumber \\
	&+ \sum_{SS^{\prime}\bm{Q}}\sum_{\bm{q}\nu}G_{S^{\prime}S}\left(\bm{Q},\bm{q}\nu\right)a^{\dagger}_{S^{\prime}\bm{Q}+\bm{q}}a_{S\bm{Q}}\left(b_{\bm{q}\nu}+b_{-\bm{q}\nu}^{\dagger}\right).
	\label{ex-ph-H}
\end{align}
On the right-hand side,	the first term approximate excitonic subsystem as a non-interacting bosonic system using the exciton creation and annihilation operators, $a_{S\bm{Q}}^{\dagger}$ and $a_{S\bm{Q}}$.
$\Omega_{S\bm{Q}}$ is the energy eigenvalue corresponding to the exicton state with the center-of-mass momentum being $\bm{Q}$ and band index being $S$.
$\Omega_{S\bm{Q}}$ can be obtained by solving the following BSE~\cite{rohlfing2000electron}:
\begin{align}
	 (E_{c\bm{k}+\bm{Q}}-E_{v\bm{k}})A^{S}_{v\bm{k},c\bm{k}+\bm{Q}}&+\sum_{v'c'\bm{k}'}K_{vc\bm{k},v^{\prime}c^{\prime}\bm{k}^{\prime}}(\bm{Q})
	   \nonumber \\
	& \times A^{S}_{v^{\prime}\bm{k}^{\prime},c^{\prime}\bm{k}^{\prime}+\bm{Q}}=\Omega_{S\bm{Q}}A^{S}_{v\bm{k},c\bm{k}+\bm{Q}}.
\label{BSE}
\end{align}
The eigen function, exciton envelope function, of Eq.~(\ref{BSE}) is the expansion coefficient of exciton wave function in the electron-hole basis,
\begin{equation}
 \chi_{S}^{\bm{Q}}(\bm{r}_{e},\bm{r}_{h})=\sum_{vc\bm{k}}A_{v\bm{k},c\bm{k}+\bm{Q}}^{S}\psi_{c\bm{k}+\bm{Q}}(\bm{r}_{e})\phi_{v\bm{k}}^{\ast}(\bm{r}_{h}),
\label{Ex-wfn}
\end{equation}
 where $\psi_{c\bm{k}+\bm{Q}}$ ($\phi_{v\bm{k}}$) is the Bloch wavefunction of the electron (hole).
The second term of Eq.~(\ref{ex-ph-H}) treats the nuclear subsystem as a non-interacting phonon system under harmonic approximation, where $b_{\bm{q}\nu}$ ($b_{\bm{q}\nu}^{\dagger}$) is the phonon annihilation (creation) operator and $\omega_{\bm{q}\nu}$ is the vibration frequency.

The third term of Eq.~(\ref{ex-ph-H}) is the first-order ExPC Hamiltonian.
In the perturbative limit, the ExPC coupling matrix element can be written as~\cite{knox1963}:
\begin{align}
	 G_{S^{\prime}S}(\bm{Q},\bm{q}\nu) &= \sum_{\bm{k}vcc^{\prime}}    A_{v\bm{k},c^{\prime}\bm{k}+\bm{Q}+\bm{q}}^{S^{\prime}\ast}A_{v\bm{k},c\bm{k}+\bm{Q}}^{S}g_{c^{\prime}c}(\bm{k}+\bm{Q},\bm{q}\nu)  \nonumber \\
	 -\sum_{\bm{k}cvv^{\prime}} &   A_{v\bm{k}-\bm{q},c\bm{k}+\bm{Q}}^{S^{\prime}\ast}A_{v^{\prime}\bm{k},c\bm{k}+\bm{Q}}^{S}g_{v^{\prime}v}(\bm{k}-\bm{q},\bm{q}\nu).
\label{ex-ph-G}
\end{align}
Here, $g_{mn}(\bm{k},\bm{q}\nu)=\langle m\bm{k}+\bm{q}|\delta_{\bm{q}\nu}V^{\text{scf}}|n\bm{k}\rangle$ is the EPC matrix element for the electron or hole of the exciton.
%
To capture the spin-flip scattering, it is crucial to include SOC when calculating the EPC matrix elements.
Specifically, the initial and final Bloch states should both be two-component spinors, and the self-consistent potential of density functional theory, $V^{\text{scf}}$, should be a two-by-two matrix in the spin subspace~\cite{von1972local,fernandez2006site}, 
\begin{equation}
    V_{\alpha\beta}^{\text{scf}} = V^{\text{ext}}_{\alpha\beta} + V^{\text{H}}_{\alpha\beta}\delta_{\alpha\beta} + V^{\text{xc}}_{\alpha\beta}.
\label{V-scf}
\end{equation}
Here, $\alpha$ and $\beta$ denotes spin coordinates. 
The external potential includes the usual electron-nucleus attractive Coulomb interaction and SOC, 
\begin{equation}
    V^{\text{ext}} = -\sum_{J}\frac{Z_{J}e^{2}}{|\bm{r}-\bm{R}_{J}|}\cdot \bm{I} + \mu_{B} \bm{B}^{\text{eff}}\cdot \bm{\sigma},
\label{V-ext}
\end{equation}
where $\sigma$ is the Pauli matrix and $\bm{B}_{\text{eff}}$ is the magnetic field operator produced by the relative motion between the electron and nucleus, $\bm{B}^{\text{eff}} = \frac{\bm{E} \times \bm{p}}{2mc^{2}}$.
For non-magnetic materials and collinear magnetization, the exchange-correlation potential is diagonal in the spin space.
So for our system, i.e. monolayer TMD, the external potential is the source for spin-flip EPC and ExPC. 
More specifically, since the first term of Eq.~\ref{V-ext} is diagonal in spin, changes to the off-diagonal terms of the second term by nuclear vibrations is crucial to make spin-flip scattering happen.
In pseudopotential-based density functional theory calculations, SOC is usually included by using fully-relativistic pseudopotentials, which give equivalent descriptions to Eq. \ref{V-ext}.

After defining the ExPC matrix elements, the phonon-mediated exciton transition rate from an exciton with band index $S$ and center-of-mass momentum $\bm{Q}$ to other exciton states is calculated as~\cite{knox1963},
\begin{align}
    \Gamma_{S\bm{Q}}=\frac{2\pi}{\hbar}\frac{1}{N_{q}}\sum_{S^{\prime}\bm{q}\nu}&|G_{S^{\prime}S}(\bm{Q},\bm{q}\nu)|^{2}
    [ (n_{\bm{q}\nu}+1) \nonumber \\
    &\times\delta(\Omega_{S\bm{Q}}-\Omega_{S^{\prime}\bm{Q}+\bm{q}}-\hbar\omega_{\bm{q}\nu}) \nonumber \\
    &+n_{\bm{q}\nu}\delta(\Omega_{S\bm{Q}}-\Omega_{S^{\prime}\bm{Q}+\bm{q}}+\hbar\omega_{\bm{q}\nu})],
    \label{time}
\end{align}
where $n_{\bm{q}\nu}$ is the Bose distribution function of phonons.
The first term in the square bracket is the phonon-emission process and the second term is the phonon-absorption process.
The scattering time is the inverse of the transition rate. 

We choose monolayer WSe$_{\text{2}}$, a typical monolayer TMD, to investigate the spin-flip exciton relaxation dynamics.
The absence of inversion symmetry and the presence of three-fold rotation symmetry endow monolayer WSe$_{\text{2}}$ with valley-selective optical properties, allowing for optical access of the valley-spin polarized excitons ~\cite{xiao2012coupled,cao2012valley}.
In addition, this material has multiple nearly degenerate valleys at K and $\Lambda$ that hold rich spin physics ~\cite{hsu2017evidence,madeo2020directly,zhu2011giant,nguyen2019visualizing}. 
All density-functional theory calculations in this work are performed using the Quantum Espresso package~\cite{giannozzi2009quantum}.
The exchange-correlation potential is treated by the Perdew-Burke-Ernzerhof functional revised for solids (PBEsol)~\cite{perdew2008restoring}.
The ionic potential is treated with the optimized norm-conserving Vanderbilt pseudopotential (ONCVPSP)~\cite{hamann2013optimized}.
The fully-relativistic ONCVPSP pseudopotential is used to describe SOC.
The phonon band structures and EPC matrix elements are calculated based on the density-functional perturbation theory~\cite{baroni2001phonons} using the Quantum Espresso package~\cite{giannozzi2009quantum}.
SOC is included for the calculations of EPC matrix elements.
The $GW$-BSE calculations are performed to obtain exciton eigenvalues and wave functions using the BerkeleyGW package~\cite{rohlfing2000electron,deslippe2012berkeleygw}.
We adopt the effective mass approximation for the exciton energies in Eq. \ref{time} to calculate the phonon-mediated exciton relaxation times.
Interested readers may refer to Ref.~\cite{zhang2021phonon} for more details of the method and calculations.

\section{Results and discussions}
\begin{figure}
	\centering
	\includegraphics[width=1.0\columnwidth]{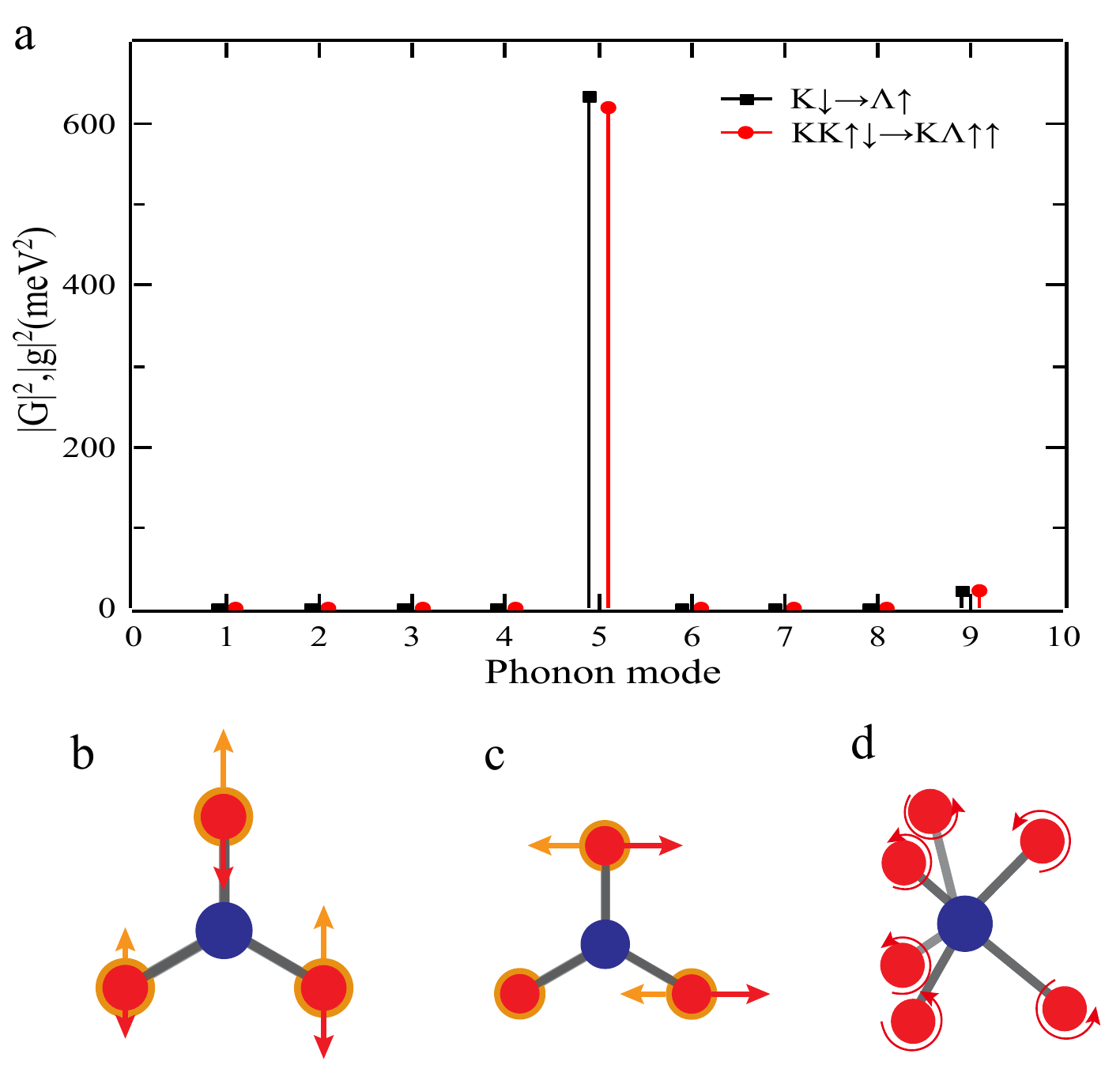}
	\caption{(a) The modulus square of calculated ExPC matrix elements, $|G|^{2}$, for the scattering from KK$\uparrow \downarrow$ 1$s$ to K$\Lambda\uparrow\uparrow$ 1$s$ and the modulus squares of the EPC matrix elements ($|g|^{2}$) for the scattering between conduction-band edges of K$\downarrow$ and $\Lambda\uparrow$. The conventions for the exciton valley-spin labels are included in the main text.
	(b) Top-view schematics of a vibration snapshot for phonon mode 4 .
	(c) and (d) are top-view and side-view schematics of a  vibration snapshot for phonon mode 5, respectively.
	W atom (Se atom) is illustrated as a blue (red or orange) ball.
	}
	\label{fig2}
\end{figure}
Due to the presence of in-plane reflection symmetry of the material, the crystal electric field that generates the effective magnetic field ($\bm{B}_{\text{eff}}$ in Eq. \ref{V-ext}) in the reference frame of the electron mainly aligns in the 2D plane.
Since the electron's momentum also lies in the 2D plane, the effective internal magnetic field for a Bloch electron is out-of-plane and opposite at its time-reversal k-pair. 
As seen in Figs.~\ref{fig1}c-d, a K (or $\Lambda$) electron has the opposite spin polarization to that of a degenerate electron at K$^{\prime}$ (or $\Lambda^{\prime}$).
Therefore, a prerequisite to spin-flip events among these valleys is to generate effective in-plane magnetic field by phonons. 
In fact, external in-plane magnetic field has been used to brighten the dark exciton in monolayer WSe$_{\text{2}}$ by coupling of the dark and bright excitons ~\cite{zhang2017magnetic}.
Using the formula of the effective magnetic field, we show that those phonon modes break the in-plane reflection and induce out-of-plane crystal electric field that plays important roles in the spin-flip EPC and ExPC.
Fig.~\ref{fig2}a shows the phonon-mode dependent spin-flip scattering strengths from K to $\Lambda$ valleys, i.e., 
KK$\uparrow \downarrow \rightarrow$ K$\Lambda\uparrow\uparrow$ exciton scattering and K$\downarrow \rightarrow$ $\Lambda\uparrow$ electron scattering, both corresponding to the electron spin flipping events but ignoring excitonic effects in the latter.
Here we label the excitons by the valley-spin configuration of their constituent electron and hole, with that of the hole on the left.
We denote the nine phonon modes by their energy-ascending order for the subsequent discussions.
We find that, the phonon mode 5 (Fig. \ref{fig2}c,d) leads to strong scattering in both electron-phonon and exciton-phonon levels of calculations, whereas other phonon modes have relatively weak scattering strength. 
By comparing the ExPC and EPC matrix elements, we conclude that the two scattering strengths are similar in size, which is consistent with the ExPC selection rules developed in Ref.~\cite{zhang2021phonon}--- under the conditions that the initial and final exciton states are both 1$s$ states, and the EPC matrix elements of a particular phonon mode is large, the ExPC strengths should be similar to the EPC one.

The variation of EPC matrix elements among different phonon modes can be first analyzed from group theory. 
The group of wave vector at K is $C_{3h}$ and at $\Lambda$ is $C_{1h}$.
Considering the transformation of out-of-plane spin under in-plane reflection $\hat{\sigma}_{h}$, the allowed phonon modes for spin-flip scattering should be anti-symmetric under $\hat{\sigma}_{h}$, such as modes 4 and 5 schematically shown in Figs.~\ref{fig2}b-d.
This mode dependence can also be understood by considering the physics picture that symmetric phonon modes under $\hat{\sigma}_{h}$ don't produce a net out-of-plane electric field.

Another striking feature in Fig.~\ref{fig2} is that very different scattering matrix elements appear even among the symmetry-allowed phonon modes.
Taking the close-in-energy modes 4 and 5 for example, although they are both symmetry-allowed based on group theory analyses, the scattering strengths of mode 4 are nearly zero while the strengths of mode 5 are $\sim$ 600 meV$^{2}$, showing a two orders of magnitude variation.
We note that the physical origin behind this observation is complicated, involving the phonon-mode dependent variation of potential (i.e., $\delta V^{\text{scf}}$), as well as the orbitals of the initial and final Bloch states.
To account for large spin-flip EPC matrix elements, the variation of potential should strongly couple the electron's orbitals of the initial states and those of the final states.
At K (K$^{\prime}$), the conduction bands are mainly contributed by the $d_{0}$ orbital of W atom ~\cite{zhu2011giant,liu2015electronic,feng2012intrinsic,kosmider2013large}.
In comparison, at $\Lambda$ ($\Lambda^{\prime}$), the conduction bands are mainly contributed by the $d_{+2}$($d_{-2}$) orbital of W atom~\cite{zhu2011giant,liu2015electronic,feng2012intrinsic,kosmider2013large}.
As such, the potential variation arising from mode 4 generates weaker coupling between the initial states at K and final states at $\Lambda$, compared with that from mode 5. 
Besides, comparing Fig.~\ref{fig2}b with Fig.~\ref{fig2}c, mode 5 can produce larger out-of-plane electric field changes than mode 4.
In Fig.~\ref{fig2}b, although the top Se atoms vibrate in opposition to the bottom Se atoms, the electric field variation felt by the electron around the W atom is largely canceled out.
In contrast, in Fig.~\ref{fig2}c, although the electric field variation produced by the top Se atoms is also canceled out, the bottom Se atoms can produce sizable out-of-plane electric field variation. 
\begin{figure}
	\centering
	\includegraphics[width=1.0\columnwidth]{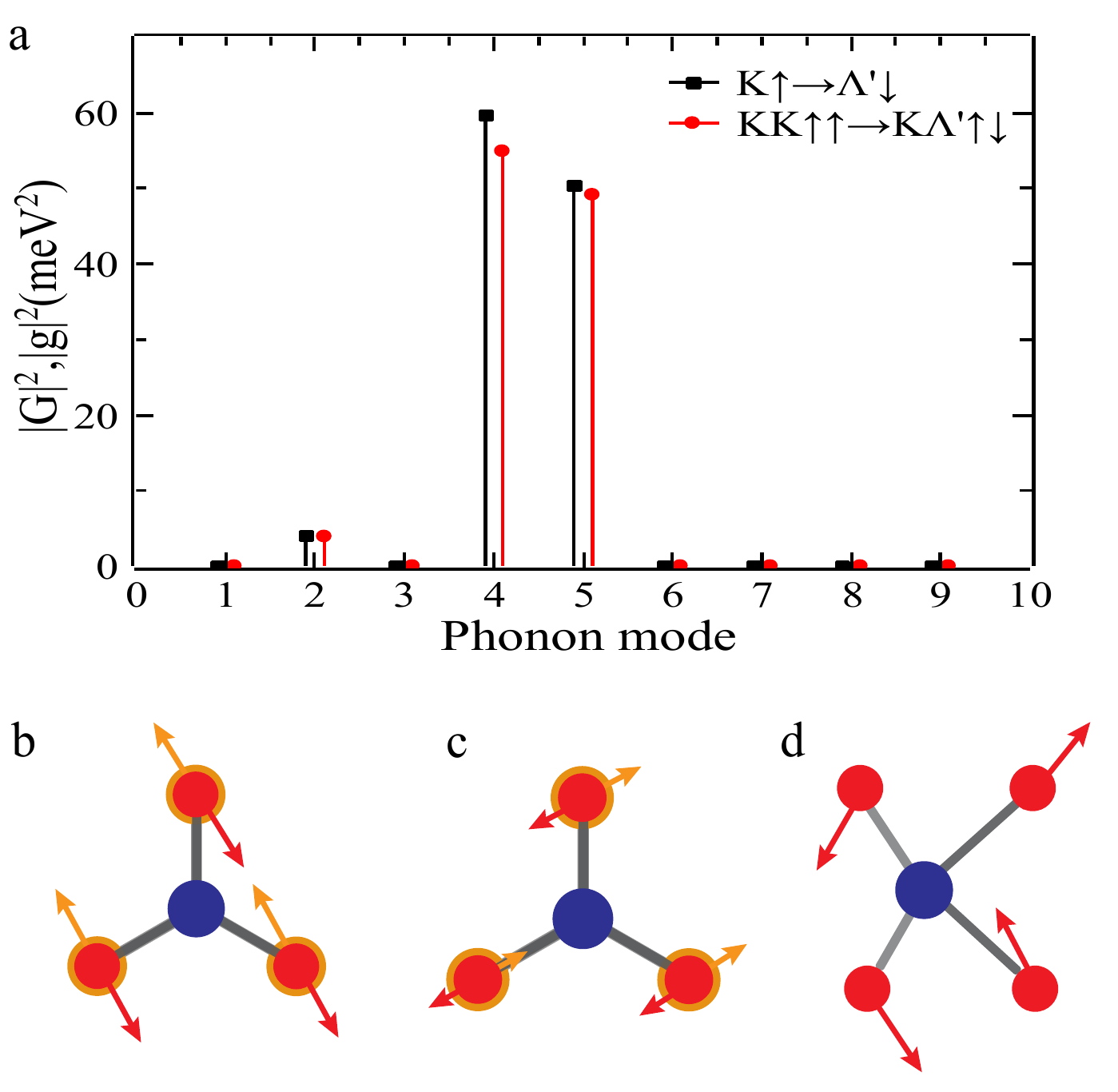}
	\caption{a) The modulus square of calculated ExPC matrix elements, $|G|^{2}$, for the scattering from KK$\uparrow \uparrow$ 1$s$ to K$\Lambda^{\prime}\uparrow\downarrow$ 1$s$ and the modulus squares of the EPC matrix elements ($|g|^{2}$) for the scattering between conduction-band edges of K$\uparrow$ and $\Lambda^{\prime}\downarrow$.
	(b) Top-view schematics of a vibration snapshot for mode 4.
	(c) and (d) are top-view and side-view schematics of a vibration snapshot for mode 5, respectively.
	}
	\label{fig3}
\end{figure}

Fig.~\ref{fig3}a shows the phonon-mode dependent spin-flip ExPC matrix elements for the scattering from KK$\uparrow \uparrow$ 1$s$ to K$\Lambda^{\prime}\uparrow\downarrow$ 1$s$ and spin-flip EPC matrix elements of K$\uparrow \rightarrow\Lambda^{\prime}\downarrow$ electron scattering.
Similar to the scattering between K and $\Lambda$ valleys, the allowed phonon modes need to be anti-symmetric under $\hat{\sigma}_{h}$, such as modes 4 and 5 schematically shown in Figs.~\ref{fig3}b-d.
At first glance, we would expect the K$\uparrow \rightarrow\Lambda^{\prime}\downarrow$ electron scattering to hold similar patterns as the K$\downarrow \rightarrow\Lambda\uparrow$ ones.
However, compared with K$\downarrow \rightarrow\Lambda\uparrow$, the spin-flip scattering strengths for K$\uparrow \rightarrow\Lambda^{\prime}\downarrow$ are significantly smaller, due to the different vibrational patterns and induced potential variation.
For example, the strengths of modes 4 and 5 are $\sim$ 60 meV$^{2}$, much smaller than that of mode 5 in Fig. \ref{fig2}.
\begin{figure}[b]
	\centering
	\includegraphics[width=0.9\columnwidth]{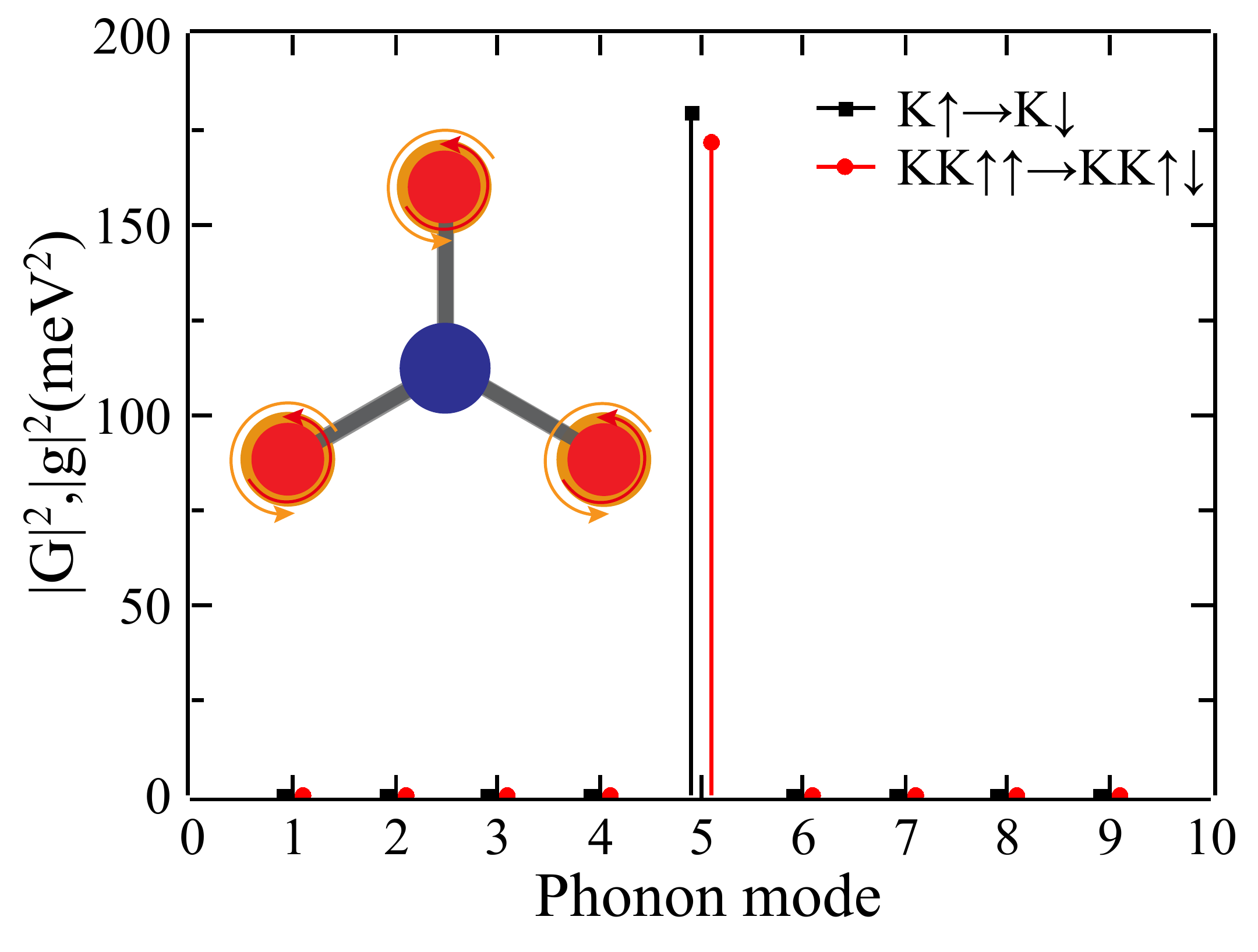}
	\caption{The modulus square of calculated ExPC matrix elements, $|G|^{2}$, for the scattering from KK$\uparrow \uparrow$ 1$s$ to KK$\uparrow\downarrow$ 1$s$ and the modulus squares of the EPC matrix elements ($|g|^{2}$) for the scattering between the two conduction-band edges at K valley. The top-view vibration schematic of $E^{\prime\prime}$ gamma-point phonon mode is shown in the inset. 
	}
	\label{fig4}
\end{figure}
\begin{figure}[b]
	\centering
	\includegraphics[width=0.9\columnwidth]{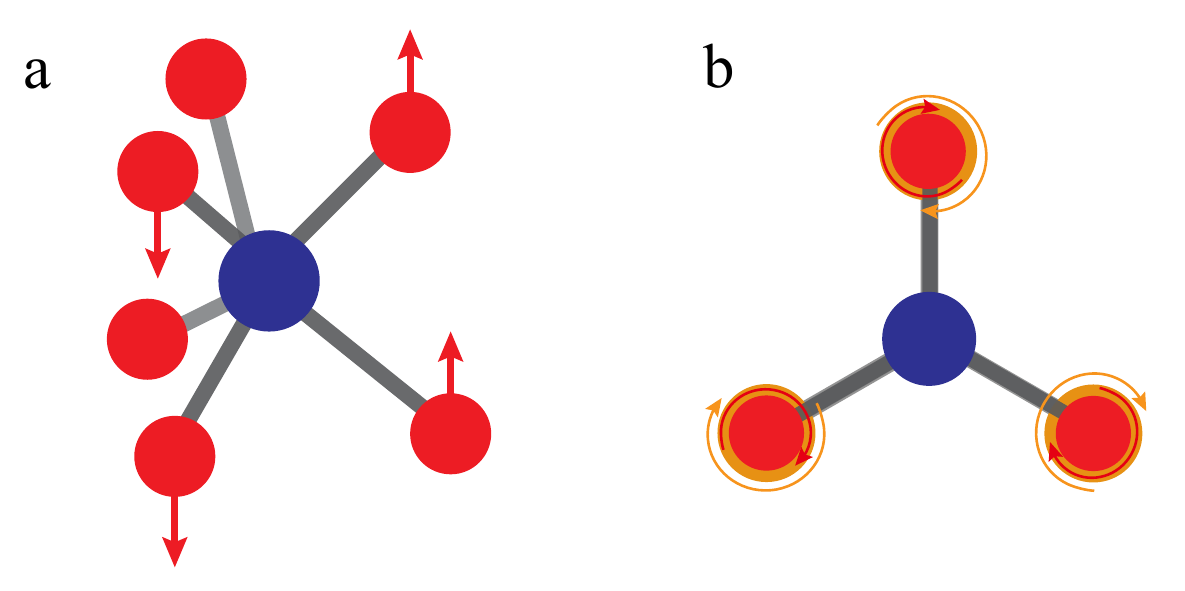}
	\caption{ Schematics of vibration snapshots for two K phonon modes which are allowed for K$\rightarrow$K$^{\prime}$ intervalley spin-flip scattering based on group theory analysis:(a) $^{2}E^{\prime\prime}$ mode scatters K$\uparrow$ to K$^{\prime}\downarrow$;(b)$A^{\prime\prime}$ mode scatters K$\downarrow$ to K$^{\prime}\uparrow$.
	}
	\label{fig5}
\end{figure}
\begin{figure}[b]
	\centering
	\includegraphics[width=0.9\columnwidth]{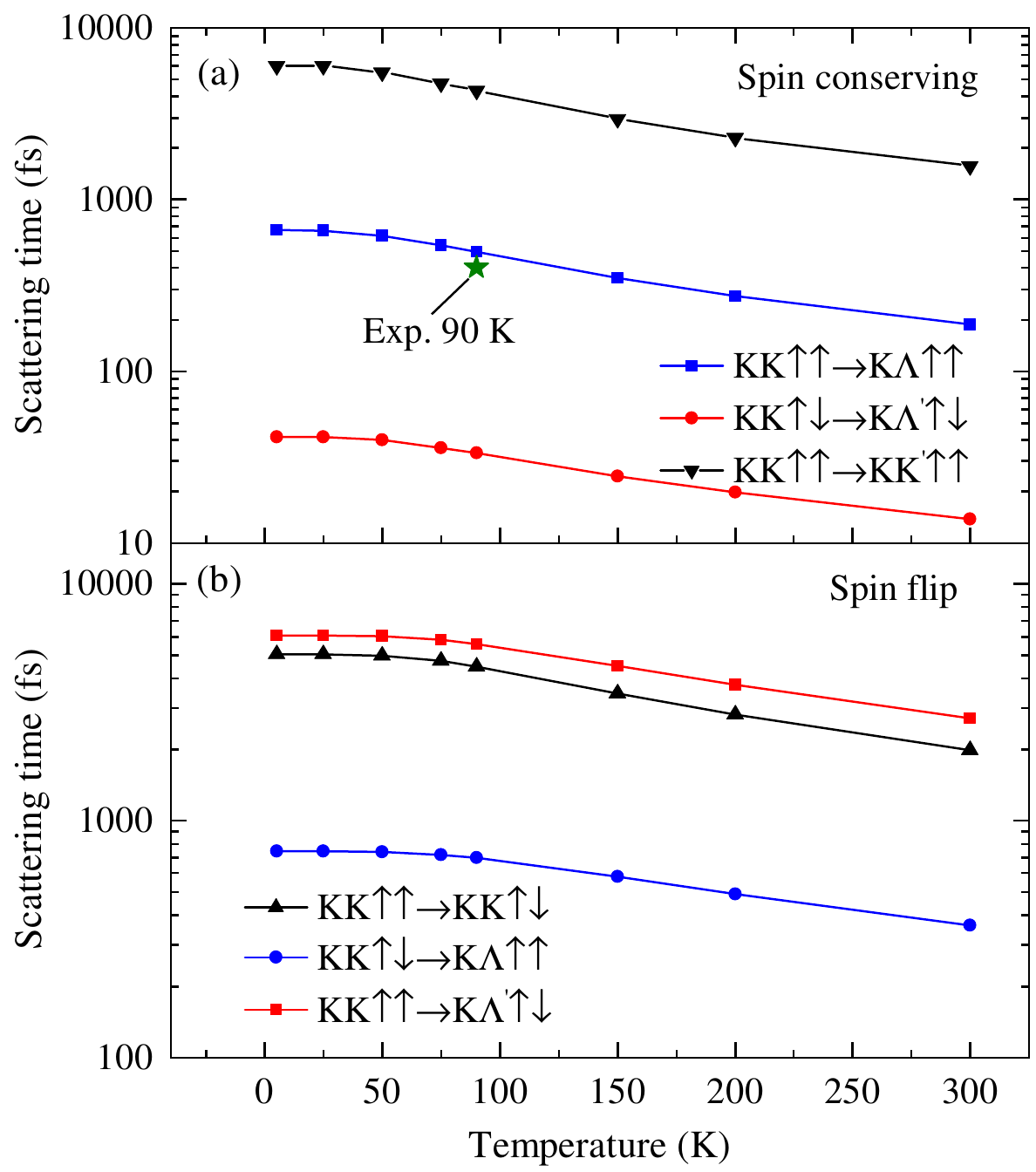}
	\caption{Calculated temperature-dependent scattering times of KK 1$s$ excitons for the (a) spin-conserving and (b) spin-flip relaxation pathways. The experimental value from Ref.~\cite{madeo2020directly} at 90 K is marked by a star.
	}
	\label{fig6}
\end{figure}

Fig.~\ref{fig4} shows the phonon-mode dependence of like-to-unlike spin exciton scattering strength, KK$\uparrow \uparrow$ 1$s$ to KK$\uparrow\downarrow$ 1$s$, and spin-flip electron scattering strengths, K$\uparrow$ to K$\downarrow$.
The allowed phonon modes are $E^{\prime\prime}$ modes under group theory analysis.
$E^{\prime\prime}$ modes have two degenerate phonon eigenvectors, which means there exists a two-component subspace that allows for arbitrary linear combinations of vibrational modes.
Under the linearly-polarized phonon vibrations, the spin-flip scattering strengths of these two modes are near 90 meV$^{2}$, which is similar to the previously calculated frozen-phonon result, $\sim$ 83 meV$^{2}$~\cite{li2019emerging}.
Under the chiral vibrations, these two modes can be transformed to one right-circularly polarized mode and one left-circularly polarized mode.
As a result, the presence of three-fold rotation symmetry requires that the right-circularly polarized can scatter K$\uparrow$ to K$\downarrow$ while the left-circularly polarized mode can scatter K$\downarrow$ to K$\uparrow$~\cite{li2019emerging,liu2019valley}.
Specifically, the scattering strengths of these two chiral phonon modes are near 180 meV$^{2}$ (twice the case of each linear polarization) or zero.
As shown in the inset of Fig.~\ref{fig4}, the circular vibration of phonon can shift the center of the Se-atom triangle and produce out-of-plane electric field. 
Note that although $E^{\prime\prime}$ phonon modes only vibrate in the plane, the spin-flip scattering strength is sizable, which is different from the case of spin-flip scattering between K and $\Lambda$ or $\Lambda^{\prime}$ valleys.
This is likely due to the fact that the initial and final states share almost the same spatial part of wavefunction, making it easier to have larger EPC matrix elements under phonon-induced potential variation \cite{li2019emerging}.
Noticing that the spin-flip scattering between K and K$^{\prime}$ valleys seems to be similar to the intra-valley scattering at K, because K phonon modes also have three-fold rotation symmetry and K$^{\prime}$ valley is also composed of $d_{0}$-orbital.
According to group theory analyses, among K phonon modes, only $^{2}E^{\prime\prime}$ mode can scatter K$\uparrow$ to K$^{\prime}\downarrow$ and only $A^{\prime\prime}$ mode can scatter K$\downarrow$ to K$^{\prime}\uparrow$.
Fig.~\ref{fig5} shows the vibration schematics of these two modes. 
However, the scattering strengths of these two modes are almost zero.
This seems to be surprising since $A^{\prime\prime} $ K phonon mode, for example, also has a similar circular vibration as $E^{\prime\prime}$ $\Gamma$ phonon mode.
The reason behind this difference lies in the destructive interference of the potential generated by K phonon modes. 
For $A^{\prime\prime} $ K phonon mode, due to the 120-degree phase difference between the motions of the three Se atoms around the W atom, the center of the Se-atom triangle actually doesn't change with time, which implies the absence of net out-of-plane electric field.
Similarly, for $^{2}E^{\prime\prime}$ K phonon mode, although the top Se atoms vibrate along the the same direction as the bottom Se atoms, the out-of-plane electric field variation is almost canceled out.

Finally, we give spin-flip exciton-phonon relaxation times as shown in Fig.~\ref{fig6}b. 
The temperature dependence is caused by the phonon occupation number (see Eq.~\ref{time}).
These spin-flip relaxation pathways are crucial to understanding the exciton relaxation dynamics of monolayer WSe$_{2}$.
Taking the KK$\uparrow\uparrow$ 1$s$ exciton as the initial state, there are mainly five inter-valley relaxation pathways, three spin-conserving: (1) KK$\uparrow \uparrow \rightarrow$K$\Lambda$ $\uparrow \uparrow$; (2) KK$\uparrow \uparrow \rightarrow$K$\Lambda^{\prime}$ $\uparrow \downarrow$; (3) KK$\uparrow \uparrow \rightarrow$KK$^{\prime}$ $\uparrow \uparrow$; and two spin-flip: (4) KK$\uparrow \uparrow \rightarrow$KK$\uparrow \downarrow \rightarrow$K$\Lambda \uparrow \uparrow$; (5) KK$\uparrow \uparrow \rightarrow$ KK$\uparrow \downarrow \rightarrow$K$\Lambda^{\prime}\uparrow \downarrow$.
Since the two spin-flip channels both involve the intra-valley spin-flip process, the comparison of scattering times between this process, KK$\uparrow \uparrow \rightarrow$KK$\uparrow \downarrow$, and the three spin-conserving ones is important for complete understanding of the scattering mechanism of the experiment in Ref.~\cite{madeo2020directly}.
Therefore, we also show the spin-conserving exciton-phonon relaxation times in Fig.~\ref{fig6}a, which has been partly discussed in Ref.~\cite{zhang2021phonon}.
Unfortunately, there have been no direct experimental measurements of spin-flip exciton relaxation times for monolayer WSe$_{2}$.
We hope our full $\textit{ab initio}$ study can stimulate more experimental studies of exciton relaxation dynamics for TMD.

\section{Conclusions}
In conclusion, we performed systematic $\textit{ab initio}$ calculations of EPC and ExPC of spin-flip scattering in monolayer WSe$_{\text{2}}$. 
Based on the mechanism of Ising-type SOC in the valleys, we propose that the effective in-plane internal magnetic field induced by phonon vibration is a prerequisite to flip spin.
In addition, we find sensitive phonon-mode dependence for the spin-flip scattering rate at both EPC and ExPC levels of calculations.
Finally, the calculated temperature-dependent exciton relaxation times can be directly compared with future experiments. 
Considering the similarity of electronic structure and phonon vibration among the TMD family, these numerical results can shed light on the microscopic origin for spin-flip exciton relaxation dynamics of other TMD systems. 

\section*{Acknowledgments}
This research was primarily supported by NSF through the University of Washington Materials Research Science and Engineering Center DMR-1719797.

\section*{References}
\bibliographystyle{iopart-num}
\providecommand{\newblock}{}

\end{document}